\documentclass[conference,a4paper]{APSIPA2021}
\usepackage{amsmath}
\usepackage{graphicx}
\usepackage{multirow}
\usepackage{booktabs}
\usepackage{threeparttable}
\usepackage[backend=biber,style=ieee,]{biblatex}
\usepackage{geometry}
\usepackage{bm}
\usepackage{fancyhdr}

\fancypagestyle{firststyle}{
  \fancyhf{}
  \fancyhead[C]{2025 Asia Pacific Signal and Information Processing Association Annual Summit and Conference (APSIPA ASC)}
}

\begin{document}

\title{End-to-End Simultaneous Dysarthric Speech Reconstruction with Frame-Level Adaptor and Multiple Wait-k Knowledge Distillation}

\author{
\authorblockN{
Minghui Wu\authorrefmark{1}, Haitao Tang\authorrefmark{2}, Jiahuan Fan\authorrefmark{2}, Ruizhi Liao\authorrefmark{2}, Yanyong Zhang\authorrefmark{1}
}

\authorblockA{
\authorrefmark{1}
University of Science and Technology of China, China
\\
E-mail: wmhsky@mail.ustc.edu.cn, yanyongz@ustc.edu.cn
}

\authorblockA{
\authorrefmark{2}
iFlytek Co., Ltd., China}


}

\maketitle
\thispagestyle{firststyle}
\pagestyle{fancy}

\begin{abstract}
Dysarthric speech reconstruction (DSR) typically employs a cascaded system that combines automatic speech recognition (ASR) and sentence-level text-to-speech (TTS) to convert dysarthric speech into normally-prosodied speech. However, dysarthric individuals often speak more slowly, leading to excessively long response times in such systems, rendering them impractical in long-speech scenarios. Cascaded DSR systems based on streaming ASR and incremental TTS can help reduce latency. However, patients with differing dysarthria severity exhibit substantial pronunciation variability for the same text, resulting in poor robustness of ASR and limiting the intelligibility of reconstructed speech. In addition, incremental TTS suffers from poor prosodic feature prediction due to a limited receptive field. In this study, we propose an end-to-end simultaneous DSR system with two key innovations: 1) A frame-level adaptor module is introduced to bridge ASR and TTS. 
By employing explicit-implicit semantic information fusion and joint module training, it enhances the error tolerance of TTS to ASR outputs.
2) A multiple wait-k autoregressive TTS module is designed to mitigate prosodic degradation via multi-view knowledge distillation. Our system has an average response time of 1.03 seconds on Tesla A100, with an average real-time factor (RTF) of 0.71.
On the UASpeech dataset, it attains a mean opinion score (MOS) of 4.67 and demonstrates a 54.25\% relative reduction in word error rate (WER) compared to the state-of-the-art.
Our demo is available at: \url{https://wflrz123.github.io/}
\end{abstract}

\section{Introduction}
\begin{figure}[t]
  \centering
  \includegraphics[width=\linewidth]{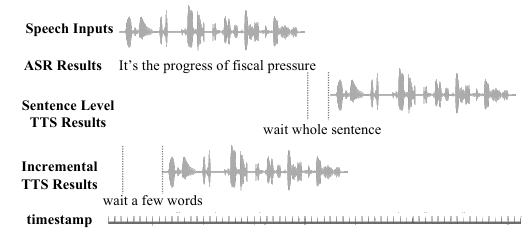}
  \vspace{-0.8cm}
  \caption{Cascaded DSR system based on ASR and TTS.}
  \vspace{-0.7cm}
  \label{fig:cascaded_dsr}
\end{figure}

Dysarthric speech reconstruction (DSR) aims to convert abnormal prosody into normal one while preserving original semantics and speaker timbre~\cite{wang2020end, wang2022speaker, wang2024unit}. A common solution involves a cascaded system~\cite{chen2024colm, fatemeh2024enhancement} that integrates automatic speech recognition (ASR)~\cite{turrisi2021deep} and sentence-level text-to-speech (TTS)~\cite{rudzicz2013adjusting}, as illustrated in Fig.~\ref{fig:cascaded_dsr}. In this approach, TTS performs rapid synthesis only after ASR has fully recognized the entire sentence. However, dysarthric patients face challenges in articulation and generally speak more slowly~\cite{wang2022speaker, liu2024two}. As a result, the reconstruction latency of sentence-level cascaded systems would significantly increase. 
This limitation hinders practical use in long-speech communication scenarios, such as meetings or telephone conversations.

To reduce sentence-level latency, simultaneous cascaded systems based on streaming ASR and incremental TTS have been introduced~\cite{sudoh2020simultaneous, barrault2023seamless} (also illustrated in Fig.~\ref{fig:cascaded_dsr}).
These systems generate reconstructed normal speech without waiting for complete output of ASR.
Streaming ASR typically employs an RNN Transducer (RNNT) model~\cite{ghodsi2020rnn, tang2023reducing}, while incremental TTS utilizes context-constrained text inputs for real-time synthesis~\cite{stephenson2021alternate, saeki2021low}.
However, dysarthric speech presents several challenges.
It is considered low-resource data~\cite{lin2023disordered} due to the difficulty in collection.
Patients with varying severity levels exhibit significant pronunciation differences for the same text~\cite{lin2023disordered}, resulting in a larger mapping space for speech-text forced alignment (FA).
Dysarthric speech ASR tasks present significant challenges, which severely limit the intelligibility of the reconstructed speech~\cite{shor2019personalizing}. 
Moreover, although incremental TTS can effectively estimate cepstral and aperiodic features, it struggles with prosodic feature prediction due to its limited receptive field~\cite{stephenson2021alternate, saeki2021low}. 
This adversely affects the quality, naturalness, and expressiveness of the reconstructed speech.

To address the aforementioned challenges, we propose an end-to-end simultaneous DSR (E2E-SDSR) system, which integrates a cascaded architecture of streaming ASR and incremental TTS. 
To mitigate the impact of ASR errors on intelligibility, an adaptor module is introduced to bridge the streaming ASR and incremental TTS. 
Given the large mapping space between dysarthric speech and text, the adaptor uses a gate-controlled excitation operator to fuse high-level implicit semantic representations from the RNNT encoder with frame-level explicit semantic information from the RNNT decoder. 
By expanding the semantic search space, this fusion mechanism enhances the error tolerance of the TTS module to ASR recognition errors. 
To improve the contextual modeling capacity of the incremental TTS, we design a multiple-wait-k autoregressive TTS module. 
This module effectively utilizes the frame-level outputs from the adaptor to process semantic frame sequences with varying lengths of $k$. Additionally, knowledge distillation (KD) based multi-task learning (MTL) is incorporated into the multiple wait-k strategy, enabling a larger receptive field model to guide a smaller one and ensuring the latter's synthesis quality approximates the former's.
In this way, the proposed E2E-SDSR system ensures both the intelligibility and natural prosody of the reconstructed speech, while maintaining real-time performance in long-speech scenarios.

\section{PROPOSED METHOD}

In this section, we provide a detailed description of the proposed system, which consists of three modules.
First, the streaming ASR module is responsible for extracting the semantic content features of dysarthric speech. 
Second, the frame-level adaptor module connects ASR and TTS, transmitting the fused semantic information from ASR to TTS. 
Third, the backend includes an incremental autoregressive TTS module based on the multiple wait-k strategy, which integrates the semantic content extracted by ASR with timbre and normal prosody to achieve the reconstruction. 
To enhance the robustness on low-resource dysarthric data, we apply two-stage training strategy:  initially, a reconstruction system is pre-trained using normal speech, followed by fine-tuning the ASR with dysarthric speech to drive the backend for the reconstruction of normal prosody.

\subsection{ASR Module Based on Conformer-RNNT}

\begin{figure}[t]
  \centering
  \includegraphics[width=\linewidth]{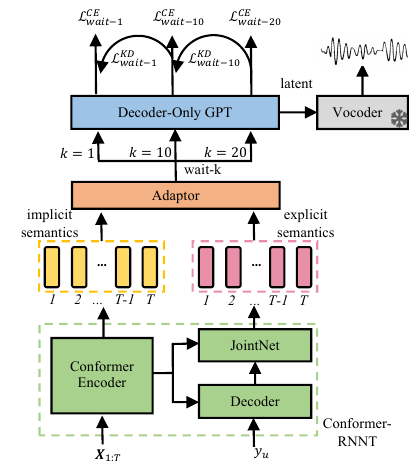}
  \vspace{-0.5cm}
  \caption{End-to-end simultaneous DSR system.}
  \vspace{-0.5cm}
  \label{fig:e2e_sdsr}
\end{figure}

To ensure the simultaneous performance of the DSR system, we use the RNNT model with inherent streaming capabilities as the ASR module. 
The RNNT model consists of an encoder, a decoder, and a joint network~\cite{graves2012sequence, sainath2020streaming}.
The encoder decouples acoustic and semantic features, with lower layers capturing speaker characteristics and higher layers representing semantic and acoustic information~\cite{tang2023reducing}. 
Conformer layers are integrated into the encoder for their global and local sequence modeling capabilities~\cite{gulati2020conformer, zhang2020transformer}. 
The decoder processes context using multi-layer LSTM networks~\cite{rao2017exploring}, and the joint network integrates encoder and decoder outputs through fully connected (FC) layers~\cite{graves2013speech} to learn explicit semantic information via the RNNT loss~\cite{huang2022training}. We present detailed steps for the Conformer-RNNT in Fig.~\ref{fig:e2e_sdsr}, which can be written as follows:

\vspace{-0.3cm}
\begin{small}
\begin{align}
\mathbf{h}_{t}^{enc} &= encoder(\mathbf{x}_{t}) \\
\mathbf{h}_{u}^{dec} &= decoder({y}_{u}) \\
\mathbf{h}_{t,u}^{joint} &= joint(\mathbf{h}_{t}^{enc}, \mathbf{h}_{u}^{dec}) \\
{Q}(\hat{y}_{u+1}|{x}_{t},{y}_{u}) &= softmax(\mathbf{h}_{t,u}^{joint})
\end{align}
\vspace{-0.2cm}
\end{small}

\noindent where acoustic feature $\mathbf{X}$ = $[\mathbf{x}_{1},\mathbf{x}_{2},\cdots,\mathbf{x}_{T}]$, previous token $\mathbf{Y}$ = $[{y}_{1},{y}_{2},\cdots,{y}_{U}]$, $\mathbf{h}_{t}^{enc}$ is the acoustic representation extracted by the acoustic encoder at the $t$-th frame, and $\mathbf{h}_{u}^{dec}$ is text representation of the $u$-th subword unit by the decoder network. 
$\mathbf{h}_{t,u}^{joint}$ is the output of the joint network. 
To reduce latency in the DSR system, the Conformer network employs a truncated context with left 64 and right 0 frames implemented via simple masks in self-attention~\cite{chen2021developing}.

\subsection{Frame-Level Adaptor Module}

Given that dysarthric speech is low-resource data with a broader speech-text mapping space compared to normal speech, the recognition performance of the RNNT model is constrained, thereby limiting the intelligibility of the reconstructed output in DSR systems.
To mitigate intelligibility degradation caused by ASR errors, an adaptor module is designed to bridge RNNT and incremental TTS, as illustrated in Fig.~\ref{fig:e2e_sdsr}.
This module is guided by the principles of informativeness and adaptability. 
We introduce semantic soft representations to enlarge the semantic search space, thereby enhancing the error tolerance of the backend TTS module. 
Specifically, the adaptor module fuses explicit semantic information $\mathbf{h}_{t,u}^{joint}$ with implicit semantic representations $\mathbf{h}_{t}^{enc}$. 
Considering the dimensional differences between these two features, we use forward operator to align $\mathbf{h}_{t,u}^{joint}$ with the frame-level $\mathbf{h}_{t}^{exp}$~\cite{graves2012sequence, sainath2020streaming}. 
$\mathbf{h}_{t}^{enc}$ primarily contains acoustic features composed of speech and silent frames~\cite{huang2023e2e}. These irrelevant silent frames are treated as semantic information during the fusion stage, increasing the search cost of semantic space. 
To address this redundancy, gate-controlled excitation operator is introduced to activate relevant frames while suppressing irrelevant ones~\cite{wu2024implicit}. 
The adaptor module is detailed as follows:

\vspace{-0.2cm}
\begin{small}
\begin{align}
\mathbf{h}_{t}^{exp} &= \begin{cases}
\mathbf{W}^{exp} \cdot \mathbf{h}_{t+1,u}^{joint} & \text{if } \hat{y}_{u+1} \text{ is } \phi, \\
\mathbf{W}^{exp} \cdot \mathbf{h}_{t+1,u+1}^{joint} & \text{else}
\end{cases} \\
\mathbf{h}_{t}^{imp} &= \mathbf{W}^{imp} \cdot \mathbf{h}_{t}^{enc} + \mathbf{b}^{imp} \\
\mathbf{h}_{t}^{apt} &= \lambda \cdot \mathbf{h}_{t}^{exp} + (1-\lambda) \cdot SwitchGLU(\mathbf{h}_{t}^{imp})
\end{align}
\vspace{-0.2cm}
\end{small}

\noindent where $\mathbf{h}_{t}^{exp}$ and $\mathbf{h}_{t}^{imp}$ represent the explicit and implicit semantic soft representations of the $t$-th frame, respectively. $\mathbf{h}_{t}^{apt}$ is the output of the adaptor module.
$\mathbf{W}^{exp}$ is trainable mapping matrix. 
$\phi$ denotes the blank symbol used in the RNNT alignment process to handle length mismatch between acoustic frames and output tokens.
$\mathbf{h}_{t}^{enc}$ is projected into the latent space through single-layer neural network. $\mathbf{W}^{imp}$ and $\mathbf{b}^{imp}$ are trainable parameters. $SwitchGLU(\cdot)$ represents switch gated linear unit~\cite{zhou2016minimal}. $\lambda$ is trainable weight of $\mathbf{h}_{t}^{exp}$ and set to 0.5 by initialization. 
The adaptor module offers two primary advantages in the DSR system: 1) Compared to using only explicit ASR outputs, it provides a broader semantic search space, thereby enhancing the tolerance of the TTS module to ASR recognition errors. 2) It supports gradient backpropagation and enables joint fine-tuning of the ASR and TTS modules, effectively reducing the performance gap between them.

\subsection{Autoregressive TTS Module Based on Multiple Wait-k}

Traditional incremental TTS systems typically rely on word-level semantic inputs~\cite{stephenson2021alternate, saeki2021low}. 
However, the ASR module requires sufficient acoustic features to decode these semantics, which often results in excessive first-frame response times in DSR systems~\cite{tang2023reducing}. 
To mitigate this issue, we introduce an autoregressive TTS model based on a decoder-only GPT architecture~\cite{casanova2024xtts}, as illustrated in Fig.~\ref{fig:e2e_sdsr}. 
It cleverly utilizes frame-level semantic outputs from the adaptor module to predict the quantized codebook entries of the reconstructed speech. 
The design of this module incorporates three key aspects: $\mathbf{h}_{t}^{apt}$ and the quantized codebook $\mathbf{c}_{t}$ share the same hop size, ensuring consistent sequence lengths and natural alignment; the quantized codebook, generated through a quantized vector quantization~\cite{betker2023better} (VQ) model, offers greater robustness compared to mel-spectrograms and has been widely adopted as a prediction target in modern TTS systems~\cite{betker2023better}; both the input and output sequences incorporate semantic and partial acoustic features, making the learning process easier.

Incremental TTS systems often suffers from poor prosody and quality in reconstructed speech due to limited receptive field. 
To overcome this, we introduce a multiple wait-k strategy based on multi-task learning into the autoregressive TTS model, enabling it to adapt to different receptive fields by various $k$ values. 
While larger $k$ values generally improve synthesis quality, they also lead to increased latency. 
To balance low response time and high synthesis quality, we integrate knowledge distillation into the multiple wait-k framework.
This allows a model with a large receptive field to guide the training of a model with a smaller receptive field, ensuring that the latter achieves synthesized speech quality closely matching that of the former.
The training objectives are defined as follows:

\vspace{-0.35cm}
\begin{small}
\begin{align}
\mathcal{L}_{wait-k}^{CE} &= CE(P(\mathbf{\hat c}_{t+1}|\mathbf{h}_{<t+k}^{apt},\mathbf{c}_{<t};\bm{\theta}),\mathbf{c}_{t+1})
\end{align}
\vspace{-0.35cm}
\end{small}

\vspace{-0.35cm}
\begin{small}
\begin{equation}
\begin{aligned}
\mathcal{L}_{wait-k}^{KD} &= D_{KL}(P(\mathbf{\hat c}_{t+1}|\mathbf{h}_{<t+k}^{apt},\mathbf{c}_{<t};\bm{\theta})\ || \\ 
&\ \ \ \ \ P(\mathbf{\hat c}_{t+1}|\mathbf{h}_{<t+k+10}^{apt},\mathbf{c}_{<t};\bm{\theta}))
\end{aligned}
\end{equation}
\vspace{-0.1cm}
\end{small}

\noindent Here, $\bm{\theta}$ represents the decoder-only GPT-based autoregressive model, which is shared across different receptive fields during multi-task learning. 
To accommodate varying receptive fields, $\mathcal{L}_{wait-k}^{CE}$ utilizes three different wait-k parameters: $k \in \{1,10,20\}$.
$\mathbf{h}_{<t+k}^{apt}$ from adaptor module and history codebooks $\mathbf{c}_{<t}$ are concatenated and fused as the input for $\bm{\theta}$. 
When $t<k$, $\mathbf{c}_{t}$ is set to fixed constant. 
$P(\mathbf{\hat c}_{t+1}|\mathbf{h}_{<t+k}^{apt},\mathbf{c}_{<t};\bm{\theta})$ represents the soft prediction output of $\bm{\theta}$. $D_{KL}(\cdot)$ is Kullback-Leibler (KL) divergence loss~\cite{van2014renyi}. 
The soft predictions from the large receptive field serve as the teacher, guiding the student model with a smaller receptive field. 
This approach enhances prosody and speech quality without introducing additional latency.
Finally, the loss function $\mathcal{L}^{TTS}$ of the incremental TTS model is composed of $\mathcal{L}_{wait-k}^{CE}$ and $\mathcal{L}_{wait-k}^{KD}$.

\vspace{-0.2cm}
\begin{small}
\begin{align}
\mathcal{L}^{CE} &=\mathcal{L}_{wait-1}^{CE} + \mathcal{L}_{wait-10}^{CE} + \mathcal{L}_{wait-20}^{CE} \\
\mathcal{L}^{KD} &= \mathcal{L}_{wait-1}^{KD} + \mathcal{L}_{wait-10}^{KD} \\
\mathcal{L}^{TTS} &= (1 - \alpha) \cdot \mathcal{L}^{CE} + \alpha \cdot \mathcal{L}^{KD}
\end{align}
\vspace{-0.2cm}
\end{small}

\noindent Here, $\alpha$ and set to 0.2 by default. To ensure simultaneous performance in entire DSR system, the latent predictions are segmented into chunks and passed to a Hifi-GAN-based vocoder~\cite{kong2020hifi} for waveform reconstruction.

\subsection{Two-Stage Training Strategy}

Since dysarthric speech is low-resource data, we employ a two-stage training strategy to enhance system robustness. 
In the first stage, we jointly train the E2E-SDSR system, including the RNNT, adaptor, and TTS, by optimizing the RNNT loss and the TTS loss to establish a reconstruction foundation for normal speech. Subsequently, the trained parameters are fixed and used to generate inputs for vocoder training, ensuring the stability of the vocoder~\cite{casanova2024xtts}.
In the second stage, considering that the pre-trained E2E-SDSR system relies on the RNNT to correct dysarthric speech semantics, the RNNT model is fine-tuned to further enhance performance about dysarthric speech. 
We use a batch of $n$ samples $\mathcal{D}=\{(\mathbf{X}_{i}, \mathbf{Y}_{i}),(\mathbf{\tilde X}_{i}, \mathbf{\tilde Y}_{i})\}^{n}_{i=1}$, where $\mathbf{X}$ is dysarthric speech and $\mathbf{\tilde X}$ is normal speech. This balanced combination ensures that the RNNT adapts to dysarthric speech while maintaining the output distribution of normal speech consistent with the backend modules.

\section{EXPERIMENTS}

\subsection{Experimental Setups}
For the dataset, both commercial dysarthric speech data and the UASpeech dataset are applied for training and testing. The commercial dataset consists of 500 hours of real-world recordings from Chinese free-talk scenarios, involving 3,000 speakers from 35 cities in China. Speech intelligibility is classified into five levels, with the level 1 being the most severe and the level 5 being nearly normal. 
Our study focuses on levels 3 and 4.
The UASpeech dataset~\cite{yu2018development} is the largest open-source English dysarthric speech dataset, comprising 29 speakers. Four speakers with varying intelligibility levels are selected for testing: M05 (middle), F04 (middle), M07 (low), and F02 (low).

For the RNNT module, we applied log-Mel filterbank as input acoustic features. The encoder employs 12-layers Conformer architecture as described in ~\cite{tang2023reducing}. 
The decoder consists of an embedding and an LSTM layer. 
For the TTS module, the decoder-only GPT model contains 15 Transformer layers~\cite{casanova2024xtts}. The VQ model follows the design in~\cite{betker2023better} and uses the same hop size of 640 ms as the RNNT encoder for 16 kHz speech. 

For evaluation metrics, the Whisper-Base model~\cite{graham2024evaluating} is applied to obtain recognition results and computes intelligibility by word error rate (WER). The 5-scale mean opinion score (MOS)~\cite{casanova2024xtts} is used to evaluate naturalness.

\subsection{Results and Analysis}
~\textbf{Adaptor Ablation.} Table~\ref{experiment_1} compares the performance of the adaptor module on the commercial dataset. 
To evaluate the improvement in tolerance to ASR errors, the ASR word-level outputs are fed directly into a sentence-level TTS module in configuration E2.
The ASR module uses streaming RNNT model, and the sentence-level TTS module employs a decoder-only GPT-based XTTS model~\cite{casanova2024xtts}. 
This combination ensures that the front-end and back-end modules are comparable to our proposed system.
In E3, the adaptor module applies only the explicit representation from Eq. (5) as incremental TTS input. In E4, it uses only the implicit representation from Eq. (6). 
Explicit and implicit representations are fused by learnable $\lambda$ in E5. 
E6 extends E5 by applying a gate-controlled excitation operator to the implicit representation. In E3$\sim$E6, multiple wait-k is fixed for 10 without multi-task learning to eliminate the influence of different receptive fields. 
We observed that both E3 and E4 improve TTS tolerance to ASR errors compared to E2. 
Due to the inclusion of acoustic features in the implicit representation, E4 achieves higher MOS scores at levels 3 and 4. 
E5 and E6 show that combining both representations yields better results than using either alone.
Introducing SwitchGLU in E6 achieves satisfactory reconstruction performance comparable to E1. 
We also analyze the Whisper-Base recognition results on the reconstructed speech of E6 and E2. 
E6 primarily improves the recognition of similar phonetic units, i.e. /d/ and /g/, /in/ and /ing/, thereby enhancing word distinction.

~\textbf{Multiple Wait-k Ablation.} Table~\ref{experiment_2} shows the impact of the multi-task wait-k strategy on incremental TTS. We primarily compare about three aspects: different wait-k hyperparameters, multi-task joint training, and knowledge distillation. E6$\sim$E8 represent the results of different receptive fields. It can be observed that larger receptive fields result in lower WER and higher MOS scores at levels 3 and 4. 
E9 adopts multi-task learning to share incremental TTS, enabling it to adapt to different receptive fields. 
E10 extends E9 by integrating knowledge distillation strategy, where the large receptive field model guides the training of the small receptive field model. Compared to E6$\sim$E8, small receptive field benefits from the prior knowledge of large one, improving recognition performance and MOS scores for reconstructed speech. For example, E10 (wait-k=0) achieves performance closer to E9 (wait-k=10).

\begin{figure}[!t] 
\centering 
\includegraphics[width=0.45\textwidth]{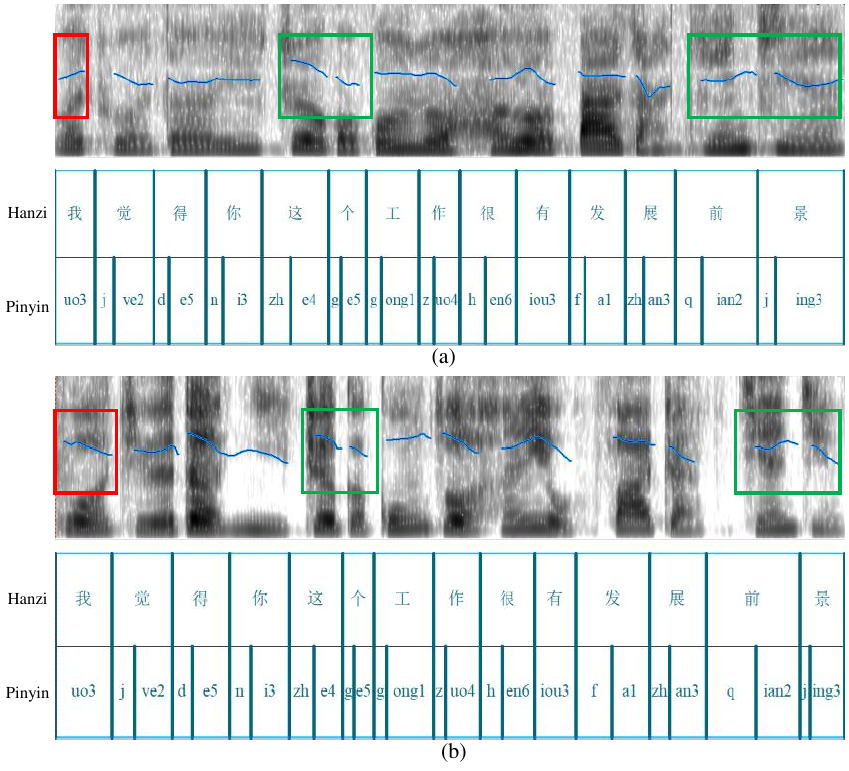} 
\caption{Comparison prosody on different receptive field for  incremental TTS . (a) is E7; (b) is E8.}
\label{analysis}
\end{figure}

We analyze the prosody on different receptive fields for incremental TTS in Fig.~\ref{analysis}. (a) is lower receptive field model, (b) is larger one. For easier understanding, Hanzi and Pinyin are used, corresponding to mel-spectrograms. It can be seen that the speech length and pauses remain unchanged in (a) and (b). In (b), prosody and naturalness are significantly improved. The incorrect tones in the red boxes are noticeably corrected, and the pathologically prolonged segments in the green boxes are reconstructed into normal one.

\begin{table}[!t]
\centering
\caption{Results of the adaptor module on commercial dataset.}
\setlength{\tabcolsep}{1.0 mm}{
\fontsize{9}{12}\selectfont
\begin{tabular}{ccccccc}
\toprule

\multirow{2}{*}{ID} & \multirow{2}{*}{Model}      & \multicolumn{2}{c}{3-rd Level} & \multicolumn{2}{c}{4-th Level} \\
                    &                             & WER            & MOS           & WER            & MOS           \\ \hline
E1                  & Original                    & 49.42          & 2.08          & 33.71          & 2.48          \\
E2                  & ASR+sentence-level TTS      & 25.93          & 3.32          & 12.91          & 4.71          \\
E3                  & Adaptor w/ Eq. (5)          & 25.63          & 3.27          & 12.89          & 4.24          \\
E4                  & Adaptor w/ Eq. (6)          & 25.07          & 3.41          & 12.23          & 4.53          \\
E5                  & Adaptor w/ Eq. (5), Eq. (6) & 23.67          & 3.64          & 11.55          & 4.76          \\
E6                  & E5 w/ SwitchGLU             & 22.58          & 3.66          & 10.84          & 4.79          \\

\bottomrule
\end{tabular}}
\label{experiment_1}
\end{table}

\begin{table}[!t]
\centering
\caption{Results of the multi-task wait-k on incremental TTS.}
\setlength{\tabcolsep}{1.0 mm}{
\fontsize{9}{12}\selectfont
\begin{tabular}{ccccccc}
\toprule

\multirow{2}{*}{ID}  & \multirow{2}{*}{Model} & \multicolumn{2}{c}{3-rd Level} & \multicolumn{2}{c}{4-th Level} \\
                     &                        & WER            & MOS           & WER            & MOS           \\ \hline
E1                   & Original               & 49.42          & 2.08          & 33.71          & 2.48          \\
E7                   & wait-k = 1             & 24.83          & 3.28          & 14.17          & 4.02          \\
E6                   & wait-k = 10            & 22.58          & 3.66          & 10.84          & 4.79          \\
E8                   & wait-k = 20            & 21.07          & 4.44          & 9.73           & 4.84          \\ \hline
\multirow{3}{*}{E9}  & wait-k = 1 w/ MTL       & 24.61          & 3.30          & 14.08          & 4.05          \\
                     & wait-k = 10 w/ MTL      & 22.58          & 3.73          & 10.84          & 4.81          \\
                     & wait-k = 20 w/ MTL      & 20.75          & 4.53          & 9.84           & 4.85          \\ \hline
\multirow{3}{*}{E10} & wait-k = 1 w/ MTL, KD   & 22.25          & 3.82          & 10.37          & 4.81          \\
                     & wait-k = 10 w/ MTL, KD  & 21.52          & 3.93          & 10.09          & 4.82          \\
                     & wait-k = 20 w/ MTL, KD  & 20.95          & 4.41          & 10.05          & 4.84         \\

\bottomrule
\end{tabular}}
\label{experiment_2}
\end{table}


\begin{table}[!t]
\centering
\caption{Performance of different DSR Systems.}
\setlength{\tabcolsep}{1.0 mm}{
\fontsize{9}{12}\selectfont
\begin{tabular}{ccccccc}
\toprule

ID & Params (M)              & FLOPS (M)               & \begin{tabular}[c]{@{}c@{}}System Response Time\\ (second)\end{tabular} & RTF  \\ \hline
E2 & 248.92                  & 400.48                  & 2.67                                                                    & 1.39 \\ \hline
E7 & \multirow{3}{*}{232.83} & \multirow{3}{*}{350.15} & 0.61                                                                    & 0.70 \\
E6 &                         &                         & 1.03                                                                    & 0.71 \\
E8 &                         &                         & 1.38                                                                    & 0.71 \\

\bottomrule
\end{tabular}}
\label{experiment_3}
\end{table}
~\textbf{Efficiency Comparison.} Table~\ref{experiment_3} compares the performance of cascaded and E2E-SDSR systems on the Tesla A100 platform. With similar parameters and FLOPS, the proposed system a faster response time and a lower RTF than the cascaded one. 
This is because our proposed system corrects and reconstructs speech simultaneously without waiting for entire sentences. Comparing E6$\sim$E8, the choice of receptive field size primarily influences the response time.

\begin{table}[!t]
\centering
\caption{The 5-scale MOS test scores and WER (\%) comparison E2E-SDSR and SOTA for UASpeech.}
\setlength{\tabcolsep}{1.0 mm}{
\fontsize{9}{12}\selectfont
\begin{tabular}{ccccccc}
\toprule

ID                   & Model                     & Metric & \begin{tabular}[c]{@{}c@{}}M05\\ (mid)\end{tabular} & \begin{tabular}[c]{@{}c@{}}F04\\ (mid)\end{tabular} & \begin{tabular}[c]{@{}c@{}}M07\\ (low)\end{tabular} & \begin{tabular}[c]{@{}c@{}}F02\\ (low)\end{tabular} \\ \hline
\multirow{2}{*}{E11} & \multirow{2}{*}{Original} & WER    & 81.70                                               & 81.70                                               & 95.60                                               & 95.90                                               \\
                     &                           & MOS    & 2.86                                                & 2.42                                                & 1.75                                                & 1.95                                                \\ \hline
\multirow{2}{*}{E12} & \multirow{2}{*}{E2E-DSR}  & WER    & 69.80                                               & 69.30                                               & 73.10                                               & 72.00                                               \\
                     &                           & MOS    & 3.61                                                & 3.47                                                & 3.72                                                & 3.87                                                \\ \hline
\multirow{2}{*}{E13} & \multirow{2}{*}{ASA-DSR}  & WER    & 62.50                                               & 76.60                                               & 62.70                                               & 65.80                                               \\
                     &                           & MOS    & 4.19                                                & 3.98                                                & 3.71                                                & 4.26                                                \\ \hline
\multirow{2}{*}{E14} & \multirow{2}{*}{Unit-DSR} & WER    & 64.40                                               & 65.50                                               & 62.10                                               & 68.30                                               \\
                     &                           & MOS    & 4.52                                                & 4.65                                                & 4.62                                                & 4.55                                                \\ \hline
\multirow{2}{*}{E15} & \multirow{2}{*}{E2E-SDSR} & WER    & 28.25                                               & 31.26                                               & 26.16                                               & 33.42                                               \\
                     &                           & MOS    & 4.65                                                & 4.72                                                & 4.68                                                & 4.65                  \\

\bottomrule
\end{tabular}}
\label{experiment_4}
\end{table}

~\textbf{SOTA Comparison.} Table~\ref{experiment_4} shows the MOS and WER comparison results on UASpeech. E11 is original speech. E12 is from~\cite{wang2022speaker}. E13 is from~\cite{wang2020end}. E14 is from~\cite{wang2024unit}. E15 is our E2E-SDSR on E10(wait-k=10). E15 achieves an average MOS of 4.67 and a 54.25\% relative reduction in average WER compared to SOTA.

\section{CONCLUSIONS}

In this work, we propose an end-to-end simultaneous DSR system that focuses on optimizing the intelligibility and the naturalness of the reconstructed speech under limited receptive fields. The system introduces two key innovations: (1) A frame-level adaptor module that mitigates for intelligibility loss caused by ASR errors through explicit-implicit information fusion and joint module training. (2) An incremental TTS module based on the wait-k strategy, which strikes a balance between low latency and improved prosody, addressing the limitations imposed by restricted receptive fields. 
Finally, the proposed method reduces the average response time by more than 50\%. 
It achieves a 54.25\% relative reduction in WER compared to SOTA on UASpeech dataset. 
On both commercial and open-source datasets, it demonstrates significant improvements in intelligibility and naturalness over original dysarthric speech.










\printbibliography

\end{document}